\pdfoutput=1 
\RequirePackage{color}
\RequirePackage{ifpdf}
\documentclass{JINST}
\usepackage{color}
\usepackage{siunitx}
\usepackage{float}

\title{Design and operation of a setup with a camera and adjustable mirror to inspect the sense-wire planes of the Time Projection Chamber inside the MicroBooNE cryostat}

\author{Benjamin Carls$^a$, Glenn Horton-Smith$^b$, Catherine C. James$^a$, Robert M. Kubinski$^a$, Stephen Pordes$^a$, Anne Schukraft$^a$\thanks{Corresponding author.}, Thomas Strauss$^c$\\
\llap{$^a$}Fermi National Accelerator Laboratory,\\
  PO Box 500, Batavia IL 60510, U.S.A.\\
\llap{$^b$}Kansas State University, Department of Physics,\\
  116 Cardwell Hall, 1228 N. 17th St., Manhattan, KS 66506-2601, U.S.A\\
\llap{$^c$}University Bern, LHEP,\\
  Sidlerstrasse 5, CH-3012 Bern, Switzerland\\
E-mail: \email{aschu@fnal.gov}}

\abstract{Detectors in particle physics, particularly when including cryogenic components, are often enclosed in vessels that do not provide any physical or visual access to the detectors themselves after installation. However, it can be desirable for experiments to visually investigate the inside of the vessel. The MicroBooNE cryostat hosts a TPC with sense-wire planes, which had to be inspected for damage such as breakage or sagging. This inspection was performed after the transportation of the vessel with the enclosed detector to its final location, but before filling with liquid argon. This paper describes an approach to view the inside of the MicroBooNE cryostat with a setup of a camera and a mirror through one of its cryogenic service nozzles. The paper describes the camera and mirror chosen for the operation, the illumination, and the mechanical structure of the setup. It explains how the system was operated and demonstrates its performance.}


\begin{document}

\section{Introduction}\label{sec:intro}

\subsection{Task description}\label{sec:task}

The MicroBooNE detector is a liquid argon time projection chamber (TPC) of rectangular geometry with dimensions 10~m (length) $\times$ 2.5~m (width) $\times$ 2.3~m (height)~\cite{uBooNEDesignReport}. A voltage of $\sim$ -128\,kV between the cathode and the anode is applied to drift electrons produced by charged particles crossing the detector to the readout anode plane. This plane is 10~m x 2.3~m wide, holding over 8,200 gold-plated stainless steel wires. The wires are strung in parallel in three different planes, where one plane has vertically oriented wires, and all three planes have a $60^{\circ}$ angle difference in orientation. The wires themselves have a 150\,\si{\micro\metre} diameter and are spaced 3~mm apart within and between the planes. The wire length varies between 15~cm and 469~cm. The wires are tensioned with 9.8\,N to create a gravitational sag of less than 0.5\,mm for a 5\,m long wire. Opposite the anode is the cathode, which is made from stainless steel sheets mounted on a support frame. Between the anode and cathode the TPC rectangular volume is enclosed by an electrical field cage, consisting of stainless steel rectangles made of 2.5~cm diameter tubing spaced 4~cm apart. Behind the readout plane, photomultiplier tubes (PMTs) are installed to allow triggering on the scintillation light of crossing particles. The scintillation light is 128\,nm (deep UV) and therefore not in the sensitive range of the PMTs. Thus, acrylic plates with wavelength shifting tetraphenyl butadiene (TPB) coatings are installed in between each of the PMTs and the anode plane. The TPB coating could be damaged if exposed to UV light while in contact with air, as shown in \cite{PMT}.\\

The TPC is housed in a cylindrical stainless steel cryostat vessel of 12~m in length and 3.8~m in diameter, which maintains the liquid argon bath surrounding the TPC during experiment data-taking. Once the TPC is installed within the cryostat and the cryostat endcap welded, no personnel access to the inside of the cryostat is possible. The fully constructed MicroBooNE TPC was transported inside the sealed (but empty) cryostat three miles across the Fermilab site from an assembly building to the detector hall. The task of viewing inside the cryostat was to confirm that there were not broken or sagged sense-wires after transportation. This would be visible through missing wires at wire attachment points along the anode frame or distortions in the wire pattern.\\

Thirty-four penetrations into the vessel provide pathways for detector readout and cryogenic plant functions; these nozzle and pipe penetrations have various geometries and are distributed primarily across the top of the cryostat. Most nozzles are less than 15~cm in diameter, and those used for detector functions are fully occupied, mainly by cables. Any device which can be fitted through a nozzle must then pass through the 1.5~cm gap between the field cage tubes in order to have a line-of-sight into the TPC interior.\\

\begin{figure}[tbp]
\centering
\includegraphics[width=\textwidth]{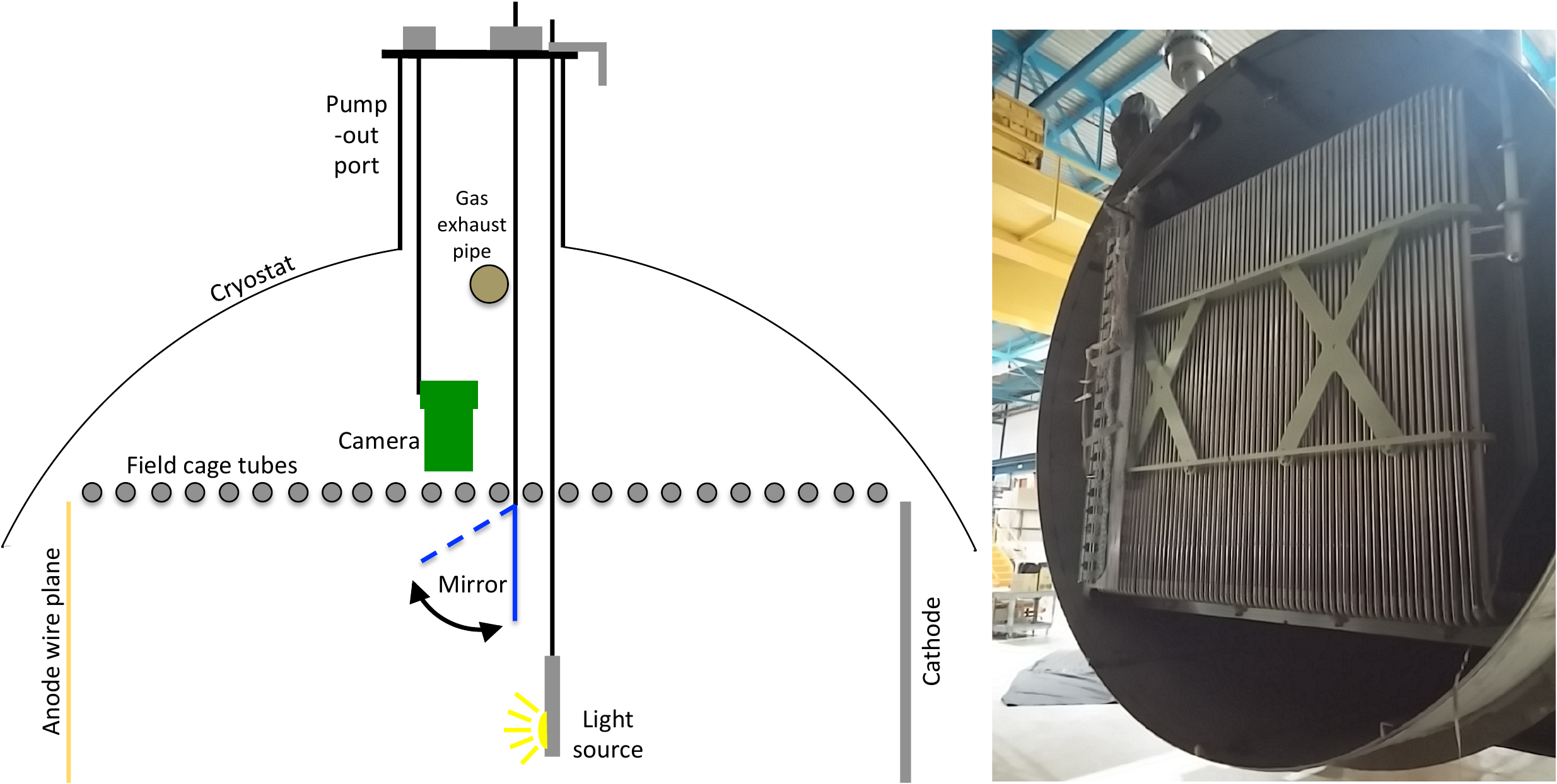}
\caption{Sketch of the setup (not to scale). Camera, mirror, and LED light source are inserted through the pump-out port. The camera rests above the field cage, while the light source and mirror slide inside the field cage. The mirror can be rotated horizontally and tilted vertically. The picture on the right shows a photograph of the TPC inside the cryostat before the endcap was welded on. The anode wires are on the left hand side, close to the cable tray. The green crosses are part of a support structure for the field cage which consists of 64 stainless steel tubes. Several nozzle penetrations are visible on the top of the cryostat. The gas exhaust pipe runs horizontally along the cryostat at the top.}
\label{fig:sketch}
\end{figure}

All of these geometric features combine to make a visual inspection of the TPC interior a challenge. This paper describes how an inspection of the TPC anode wires after closure of the cryostat, but before filling with liquid argon, was realized. A system consisting of a camera, an adjustable mirror, and a light source was designed and inserted into the cryostat through one of the cryogenic service nozzles. This system is shown schematically in Fig.~\ref{fig:sketch}. While the camera rested in a position just above the field cage, the light source and the mirror were slid inside the cage through gaps between the field cage tubes. By rotating and tilting the mirror, different areas of the interior of the TPC, in particular the anode wires, can be viewed by the camera.\\

The following subsection describes in more detail the geometry of the detector, cryostat, and the nozzle utilized for viewing the interior. It further describes the constraints this geometry imposes on the optical setup. Section~\ref{sec:setup} introduces the camera, mirror, and light source components, as well as their mechanical mounting for the setup. Section~\ref{sec:ops} describes how the optical system is operated and shows example images.\\

\subsection{Geometrical conditions}\label{sec:geom}

\begin{figure}[tbp]
\centering
\includegraphics[width=\textwidth]{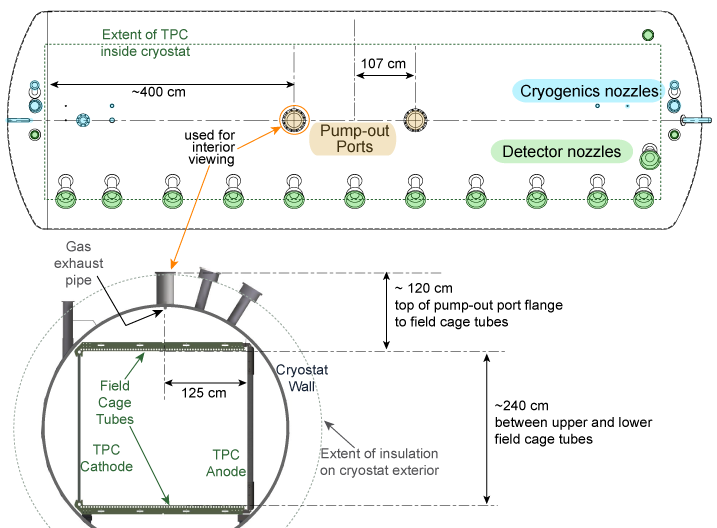}
\caption{The cryostat viewed from above and from the side. The location of the pump-out port used to insert the camera and mirror setup is marked, as well as various relevant distances.}
\label{fig:top}
\end{figure}

The largest nozzle penetrations are two ports with 27.3~cm inner diameters on the top of the MicroBooNE cryostat, each located 107~cm on either side of the center line (see Fig.~\ref{fig:top}). These nozzles were designed for pressure testing the cryostat vessel and potentially for vacuum pump-out, although this function was not utilized. As such, these pump-out ports do not contain cabling or cryogenic system monitors. The nozzles extend 53.3~cm above the vessel surface; this space between the vessel outer surface and the tops of nozzles is filled with insulation (see Fig.~\ref{fig:top}). The distance between the pump-out port's entrance at the cryostat wall and the top of the field cage tubes is 66.0~cm. Centered just below the pump-out port inside the cryostat is a gas exhaust pipe, 5.1~cm in diameter, running along the long axis of the cryostat; this pipe limited the size of objects which could be lowered through the pump-out port and into the cryostat volume.\\ 

As mentioned, the field cage consists of an array of stainless steel tubes 2.5~cm in diameter and separated by gaps of 1.5~cm. Any component entering the TPC inner volume, i.e.~the mirror and the light source, must be be thinner than the tube-to-tube spacing.\\

\begin{figure}[b!]
\centering
\includegraphics[width=\textwidth]{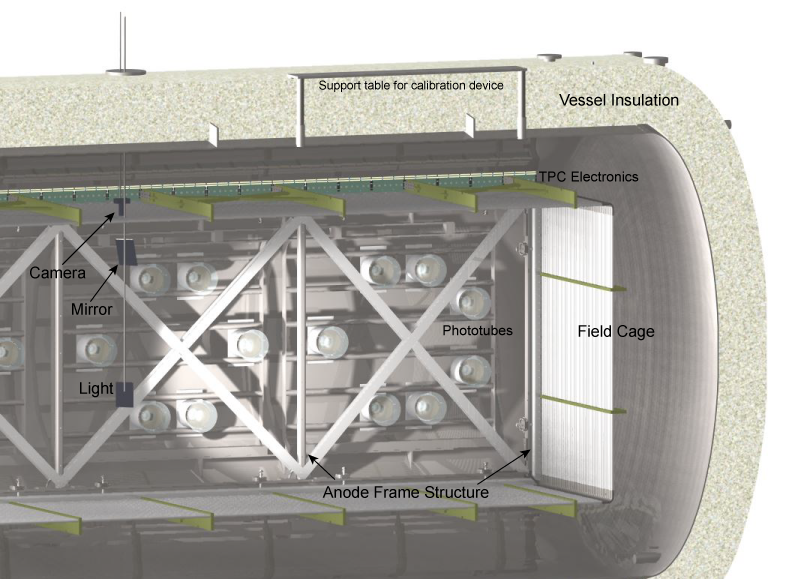}
\caption{A diagram of the vessel interior, viewed from the cathode side, showing the overall arrangement, to scale. The wires are strung on the anode frame (not visible in this picture).}
\label{fig:cutaway}
\end{figure}

With the mirror just below the top of the field cage (see Fig.~\ref{fig:sketch}), the distance to the wires directly opposite is 1.30~m (top of the anode wires) and 2.60~m (bottom of the anode wires). The furthest wires to be viewed are located in the bottom corner of the anode plane, which corresponds to a distance of almost 5~m. These distances imposed optical quality requirements on the camera and mirror to resolve individual anode wires.\\

The constraint of the pump-out port width and field cage tube spacing excluded other options to view the inside of the cryostat. Standard cameras are too large to fit between the field cage tubes and it is therefore not possible to view the wires through a direct line-of-sight or from close proximity.

A borescope was investigated and excluded on the grounds of being a potential danger to the anode wires. 
The TPC interior volume is empty space, requiring the flexible borescope tube to be supported by a rigid structure for direction to the locations of interest. 
While a support, such as a stainless steel tube, can be fitted through the described geometry, the scope lens must approach within a few cm of the wires in order to resolve them. The support tube must then be cantilevered out over the necessary distance which in turn introduces an undesirable flexibility. This results in "bounce" and vibrations as the tube is moved into position to bring the scope lens close to the wires. It was feared that the "bounce" could cause the borescope to hit and damage the anode wires. For this reason the method was excluded.\\

\section{Setup}\label{sec:setup}

Access to the inside of the MicroBooNE cryostat after closure is limited to the available service nozzels, as discussed above. Equipment that could be brought inside the vessel in order to visually inspect the sense-wires, is subject to the geometrical constraints described above. As shown in Figure \ref{fig:sketch}, the camera is inserted through the service nozzle but rests above the field cage due to its size. Therefore, a mirror that allows different tilt and rotation angles with respect to the camera optical axis is required to view the full area of the anode plane. The mirror and its tilting mechanics need to be thin enough to enter the field cage through one of the 1.5\,cm wide gaps between the tubes. As there is no light source inside the cryostat, an external light source must be inserted through the pump-out port and the field cage to illuminate the field of view of the camera, which requires a thickness of the light source apparatus of less than 1.5\,cm. Different components were tested in order to select the most suitable mirrors, light sources and mounting options for this setup. These are further described below.\\

\subsection{Camera}\label{sec:camera}

The wires align in perfect triangles. These shapes would get distorted by a sagging or broken wire. The center of the triangle is about 1\,mm from all wires, so the minimal required resolution is 0.5\,mm over a distance of 5\,m, or $1\times 10^{-4}$\,rad angular resolution, corresponding to an aperture of better than 1\,cm at visible wavelengths assuming perfect optics.\\

Another requirement was the remote operation of the camera due to inaccessibility. After examining several camera options, a Sony Alpha A6000 fitted with a Sony E 55-210~mm F4.5-6.3 OSS E-mount Zoom Lens was selected~\cite{ref:camera_spec}. Size and performance were the main factors in making these choices.
The camera's small size of 120~mm $\times$ 67~mm $\times$ 45~mm and the absence of a view finder at the top of the camera body allowed it to fit inside the pump-out port and past the gas exhaust pipe. The aperture and optical quality of the lens were sufficient to avoid any effects of diffraction or astigmatism on the image, and the focal length range of the lens allowed hundreds of wires to be seen in a single image on the camera's 24.3~megapixel sensor from up to 5~m away. The resolution achieved with this camera at 5\,m distance is approx.~0.1\,mm, and smaller for shorter distances.\\

\begin{figure}[tbp]
\centering
\includegraphics[width=\textwidth]{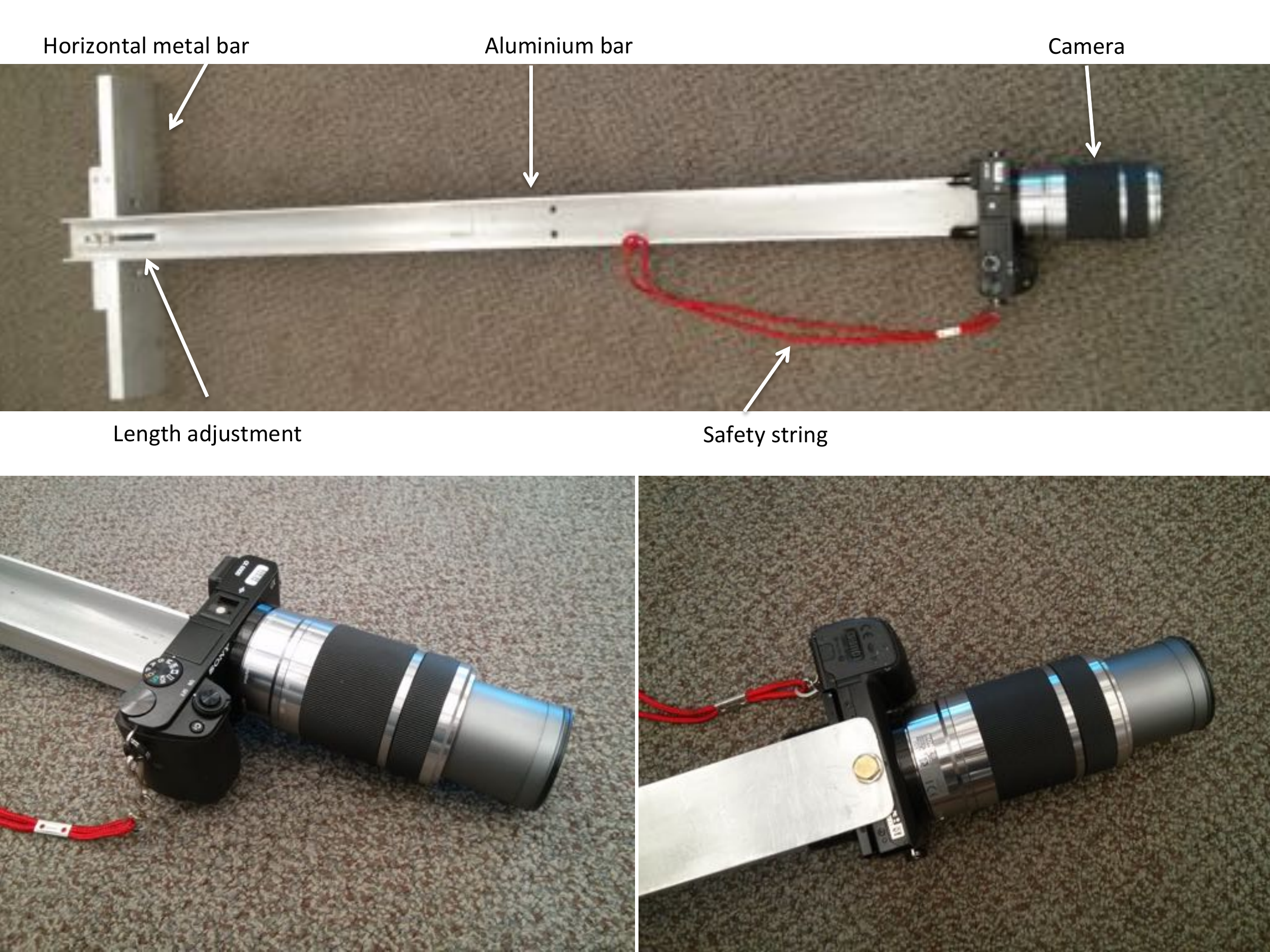}
\caption{Top: Camera apparatus. The short metal bar rests on the open flange. The two screws at the top allow a small length adjustment to correct the position according to the zoom extension of the lens. Bottom left: The camera mounted on the aluminium bar. The digital viewfinder of the camera is visible from outside when looking down to the camera inside the port. Velcro tape between camera and bar is added to protect the camera from scratching. Bottom right: The camera is attached to the aluminium bar using the camera tripod mount.}
\label{fig:camera}
\end{figure}

The camera is mounted to a T-shaped aluminium structure shown in Fig.~\ref{fig:camera}. With the camera mounted, the entire apparatus is about 1.20~m long. The short top of the T rests on the top flange of the open pump-out port. The camera is attached to the long bar using its threaded tripod mount. The T-bar can be positioned to align the center of the camera lens with a gap between a pair of field cage tubes.\\

As mentioned, when inserted, the camera lens is just above the field cage, placing it in a location where manual control of the device is impossible. The camera is therefore remotely controlled using a Google Nexus 7 (2013) Android tablet computer running the Sony PlayMemories Mobile application~\cite{ref:tablet}. The camera generated a Wi-Fi signal for the tablet to connect to. The tablet provided live images from the camera, allowing precise orientation of the camera and mirror setup. Furthermore, the camera control application allowed autofocus to be applied by touching a region on the live camera image as it appeared on the tablet's screen.\\

\subsection{Adjustable mirror}\label{sec:mirror}

\begin{figure}[tbp]
\centering
\includegraphics[width=\textwidth]{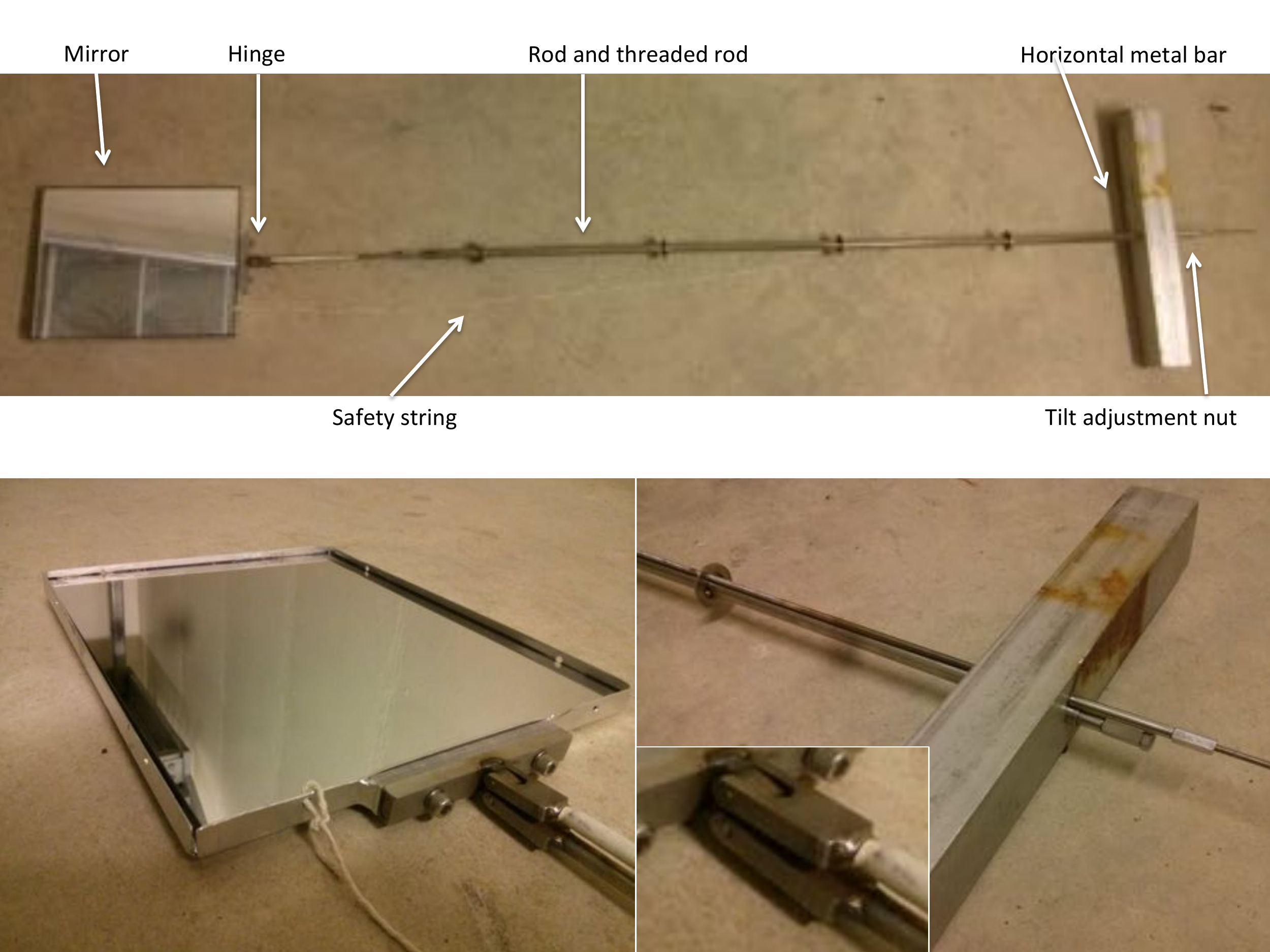}
\caption{Top: The 183~cm long mirror apparatus. Bottom left: The front surface mirror in its aluminium frame with the hinge for the tilting mechanism. The picture in the bottom center is showing a closeup of the hinge. Bottom right: The top horizontal metal bar resting on the flange with the two parallel rods sticking out of it. The longer threaded rod has a nut on it, whose rotation tilts the mirror surface.}
\label{fig:mirror}
\end{figure}

The mirror is a 20.3~cm$\times$ 25.4~cm first surface mirror made of glass~\cite{ref:mirroredmund}. The thickness is 0.6~cm, which easily fits in between a pair of field cage tubes. The mirror is glued into a frame formed from an aluminium sheet. A set of nylon screws placed on the edges of the frame and in front of the mirror prevent it from falling out should the glue fail. The mirror and the support apparatus it's attached to is shown in Fig.~\ref{fig:mirror}. The support apparatus is composed of a 183~cm long rod attached at a right angle to a wide metal bar which sits across the top of the pump-out port flange. A hinge at the top of the mirror frame allows the mirror to be tilted. A second threaded rod running parallel to the support rod is attached to the back of the mirror frame; turning a nut on this rod at the top of the apparatus adjusts the tilt angle. In order to view wires at both, the bottom and the top of the anode plane, the required angular range of the tilt is 14$^{\circ}$ to 45$^{\circ}$ with respect to the vertical mirror position. Rotation of the mirror is achieved by rotating the position of the metal bar where it rests on the pump-out flange.\\

Since viewing the mirror apparatus from outside was difficult, mirror orientations were determined using two methods. The mirror tilt angle was roughly measured by counting the number of turns the tilt adjustment nut was rotated by. A dial placed on the flange surface provided horizontal orientation of the mirror.\\

Several other mirror options were tested, but did not provide the required image quality. Two of these were types of glassless mirrors. One of the glassless mirrors was a simple polished stainless steel plate. The other was a mirror from Mirrorlite\textregistered \cite{ref:mirrorlite}, consisting of a reflective film stretched over a lightweight aluminium frame. These mirrors are shatterproof and avoid the risk of broken glass inside the cryostat. However, despite good reflectivity, the tests showed that the surfaces of these glassless mirrors were not flat enough to obtain focused images of the anode wires. A standard second-surface glass mirror was also tested, but this showed a double image due to partial reflections off of the front surface and thus was also not suitable for inspection of the wire grid structure. The risk of accidentally dropping and shattering a glass mirror when it was inside the cryostat was reduced by use of a safety line.\\

\subsection{Light source}\label{sec:lightsource}

The light source had four main requirements. First, like the mirror, it needed to fit between the field cage tubes. Second, it needed to be rotatable about the vertical axis and have an adjustable vertical position to provide maximum illumination of any region the camera was viewing. Third, the brightness had to be high enough to resolve individual wires at a distance of up to 5~m. Finally, the light source was required to be non-destructive to the UV sensitive coating of TPB applied to the wavelength shifting plates of the PMT system~\cite{PMT}, e.g. with no emission below 400\,nm wavelength. In addition, the light source had to have a minimal risk of damaging the TPC. This was achieved by minimizing the number of individual parts composing the light source setup and stress testing it to demonstrate robustness.\\

\begin{figure}[tbp]
\centering
\includegraphics[width=\textwidth]{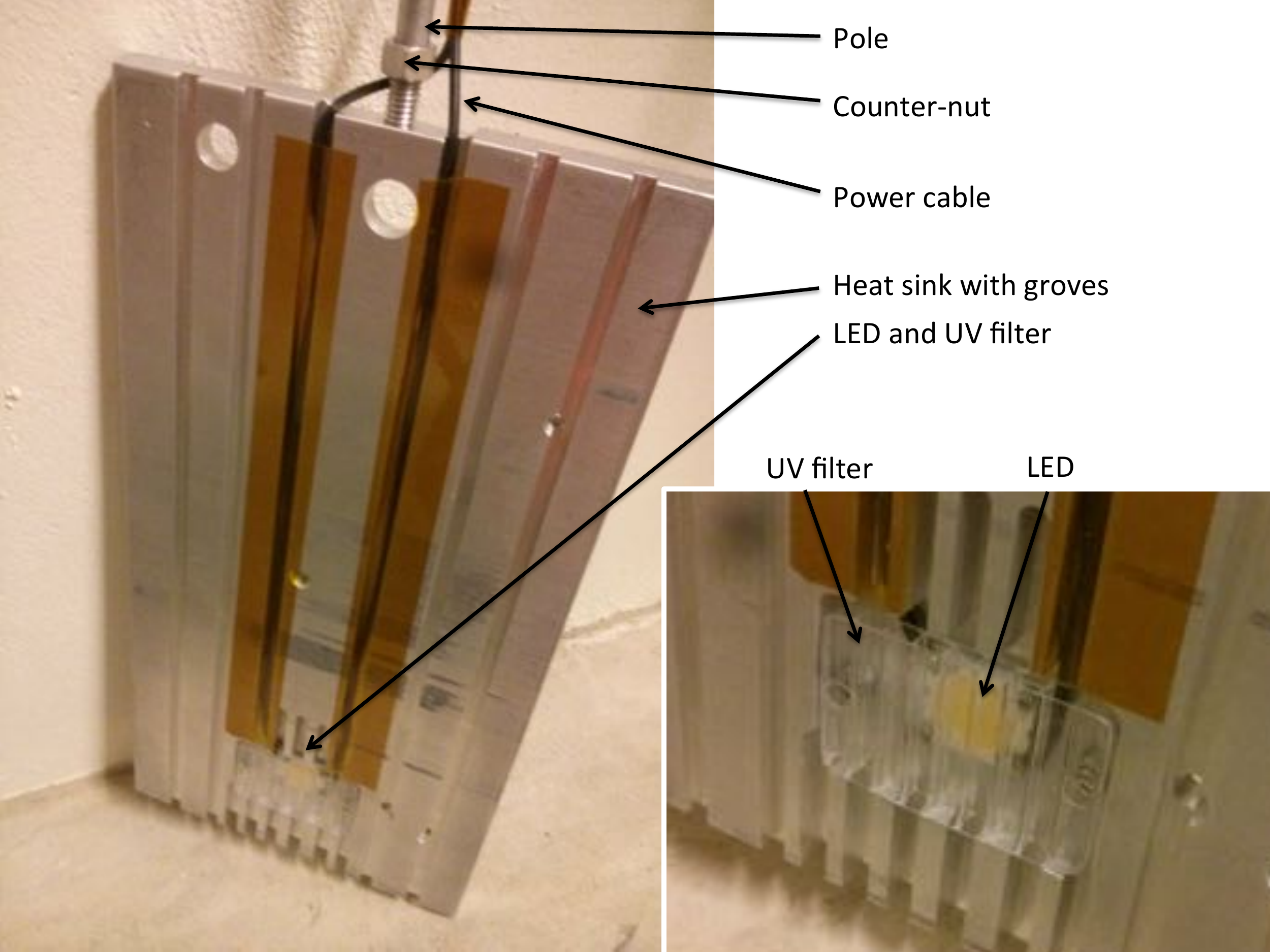}
\caption{The LED mounted to its heat sink plate and covered with the UV filter.}
\label{fig:LED}
\end{figure}

After several tests to find the optimal light source, an LED solution was selected based on the compactness of the source, the spectrum provided, and the brightness level. The LED was mounted onto a heat sink and placed behind a UV filter~\cite{UVfilter} in such a way that it would be able to fit between adjacent field cage tubes. The heat sink with LED and UV filter is attached to a long rod for insertion into the TPC interior. Raising, lowering, and rotating the mounting rod adjusts the position and direction of the light to match the viewing angle of the camera. The light source shared the same port as the camera mount and mirror as tests showed that this arrangement allowed the most light to be reflected off the wires and towards the camera. The heat sink with the mounted LED is shown in the Fig.~\ref{fig:LED}. The LED used was the Bridgelux BXRA-40E1350-B~\cite{LED}, having a thermal power of 15~W and an UV filter mounted on top to block UV light. According to the manufacturer, a well ventilated heat sink needs 65~cm$^2$ of surface area per 1~Watt of thermal power. Thus, the minimal required heat sink area for non-forced air cooling is 968~cm$^2$ for the LED type. Inside the cryostat there is virtually no air flow, thus a conservative choice was made and a larger surface area of 1,008~cm$^2$ (17.8~cm $\times$ 30.5~cm) was used. Grooves were added to the heat sink to further increase its surface area. Around the LED mounting area, the filter further obstructed air flow for cooling. To compensate, additional grooves were added to the heat sink in the area around the LED. The heat sink was made of aluminium with all grooves being vertical. During operation, the heat sink became lukewarm.\\

\begin{figure}[tbp]
\centering
\includegraphics[width=0.9\textwidth]{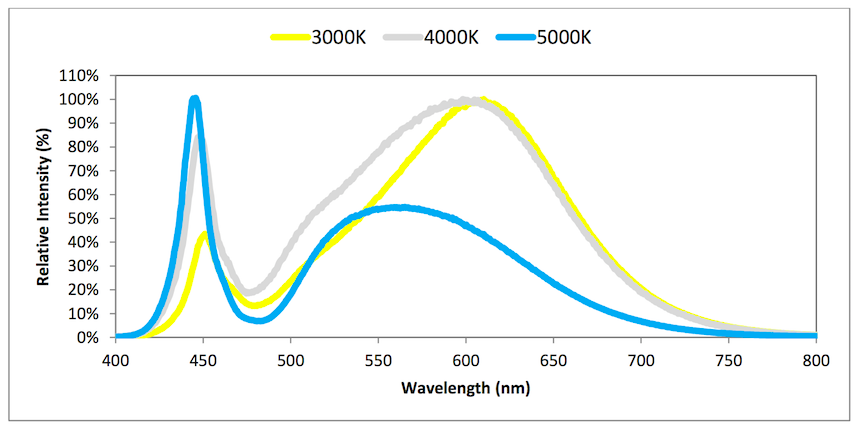}
\caption{LED light emission color spectra measured at rated current and 25$^{\circ}$C~\cite{LED}. The spectrum of the Bridgelux BXRA-40E1350-B used in this setup corresponds to the 4000K line (gray).}
\label{fig:LEDspectrum}
\end{figure}

Fig.~\ref{fig:LEDspectrum} shows the color spectrum for the selected LED (4000K line). The spectrum has a small extension into the UV region which the aforementioned filter blocked. The filter was composed of a polycarbonate material which had been shown in previous PMT system tests to prevent damage to the TPB coated plates~\cite{PMT}. As shown in \cite{UVfilter} the filter only transmits wavelengths between 400 and 2000\,nm. The material is a 0.64~cm thick double-layer panel typically used in greenhouse construction. A small piece of this material was mounted over the LED such that all the emitted light passes through and is filtered, seen in the close-up in Fig.~\ref{fig:LED}. For greenhouses, the channels formed by the material's double layer provides insulation, however in this application the channels provided cooling air flow. Long term testing of the assembly showed that the heat sink temperature did not rise above hand-warm levels and the brightness was high enough to illuminate the whole TPC in a test setup.\\

\section{Operation and performance}\label{sec:ops}

\begin{figure}[tbp]
\centering
\includegraphics[width=0.87\textwidth]{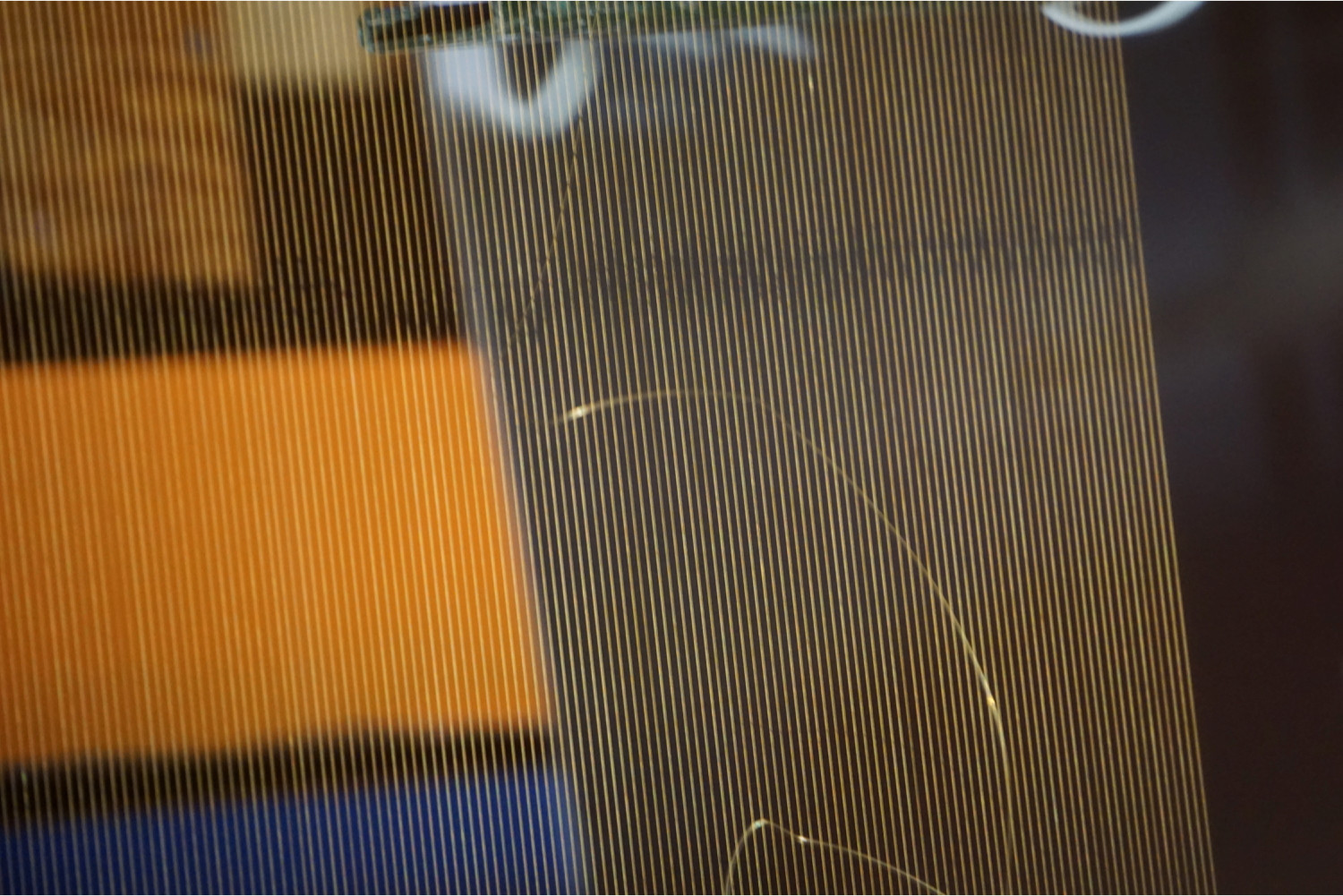}
\caption{Photograph of a broken wire on the test setup. The wire was cut at its bottom attachment point and can easily be spotted because it is curling up and distorting the wire pattern. Note that only one plane of vertical wires is strung in this particular test.}
\label{fig:153}
\end{figure}

A test setup of wires under similar illumination and field of view conditions outside of the cryostat demonstrated that broken wires would be obvious. An example image of a broken wire on the test wire pattern is shown in Fig.~\ref{fig:153}. Sagging wires become evident by non-uniformities in the wire spacing, which at the maximum viewing distance in the test setup were also shown to be easily observable.\\

For operation inside the cryostat, the camera, mirror, and light source are inserted inside the cryostat through the pump-out port as illustrated in Fig.~\ref{fig:sketch}. All support bars rest on the flange, which is covered with a protective plastic ring. The support bars, being wider than the flange opening, guarantee that none of the devices can fall inside the cryostat. Tools needed to operate the assembly, such as wrenches for adjustment nuts and flashlights, are attached to safety strings to reduce the risk of them falling into the cryostat.\\

As described in section \ref{sec:camera}, the view of the camera was followed remotely using a tablet computer. The wireless connection to the camera located just inside the cryostat was shown to be reliable. The live view on the tablet allows the blind adjustment of the mirror and light source. Orientation inside the cryostat is attained by studying the background seen behind the wires and comparing this to CAD drawings as necessary.\\

\begin{figure}[H]
\centering
\includegraphics[width=0.87\textwidth]{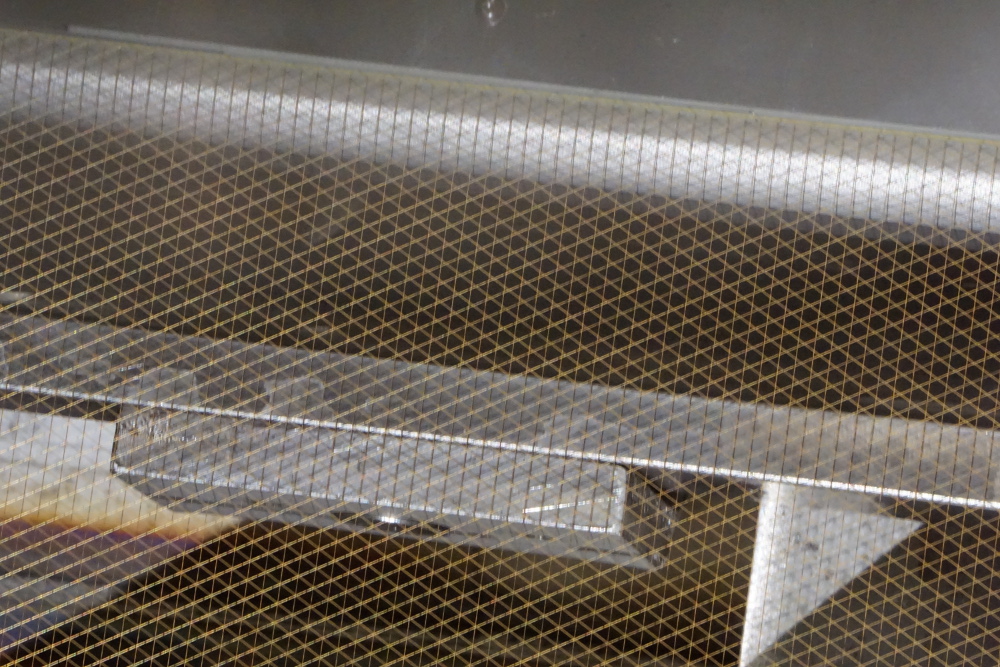}
\caption{View of the top of the anode wire plane adjacent to the pump-out port. This is the point of closest approach between mirror and object.}
\label{fig:460}
\end{figure}

\begin{figure}[H]
\centering
\includegraphics[width=0.87\textwidth]{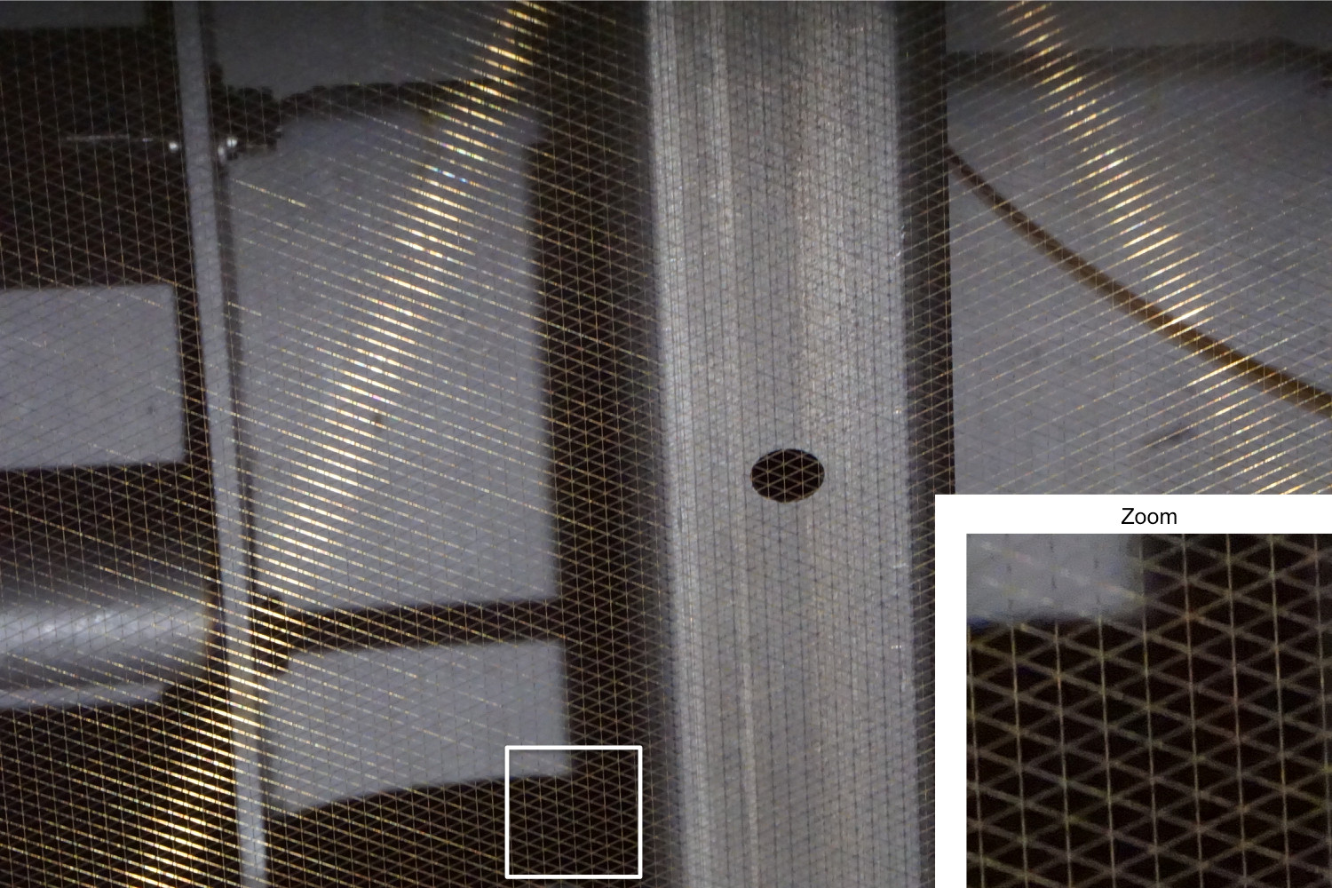}
\caption{View of a central part of the anode wire plane adjacent to the pump-out port. Support structure of the anode frame is seen in the background. The picture in the bottom right corner is a zoom of the white square in the large picture.}
\label{fig:437}
\end{figure}

\begin{figure}[H]
\centering
\includegraphics[width=0.87\textwidth]{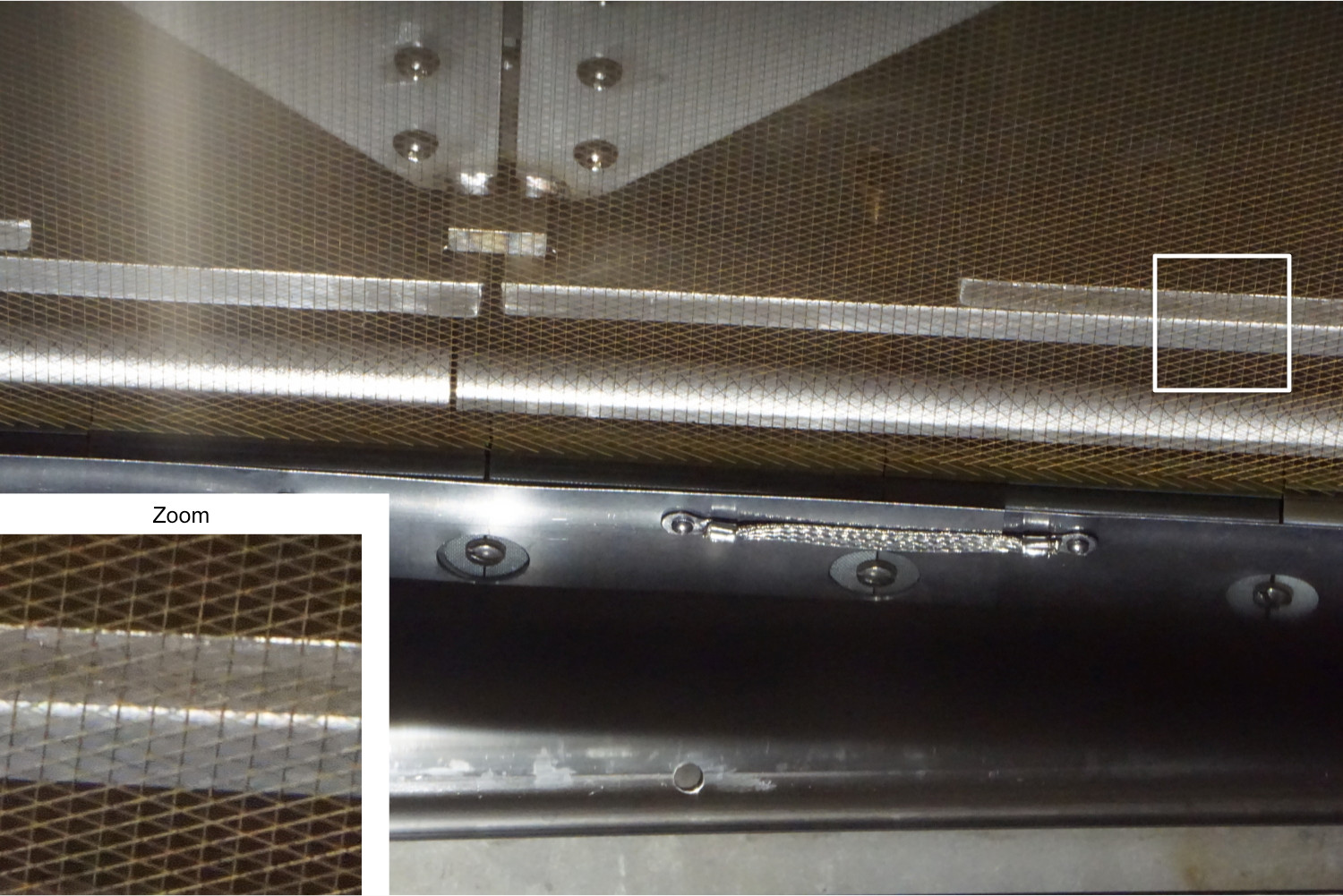}
\caption{View of the bottom of the anode wire plane adjacent to the pump-out port. The distance between the mirror and the wire plane is about 2.60~m. The picture in the bottom left corner is a zoom of the white square in the large picture.}
\label{fig:451}
\end{figure}

\begin{figure}[H]
\centering
\includegraphics[width=0.87\textwidth]{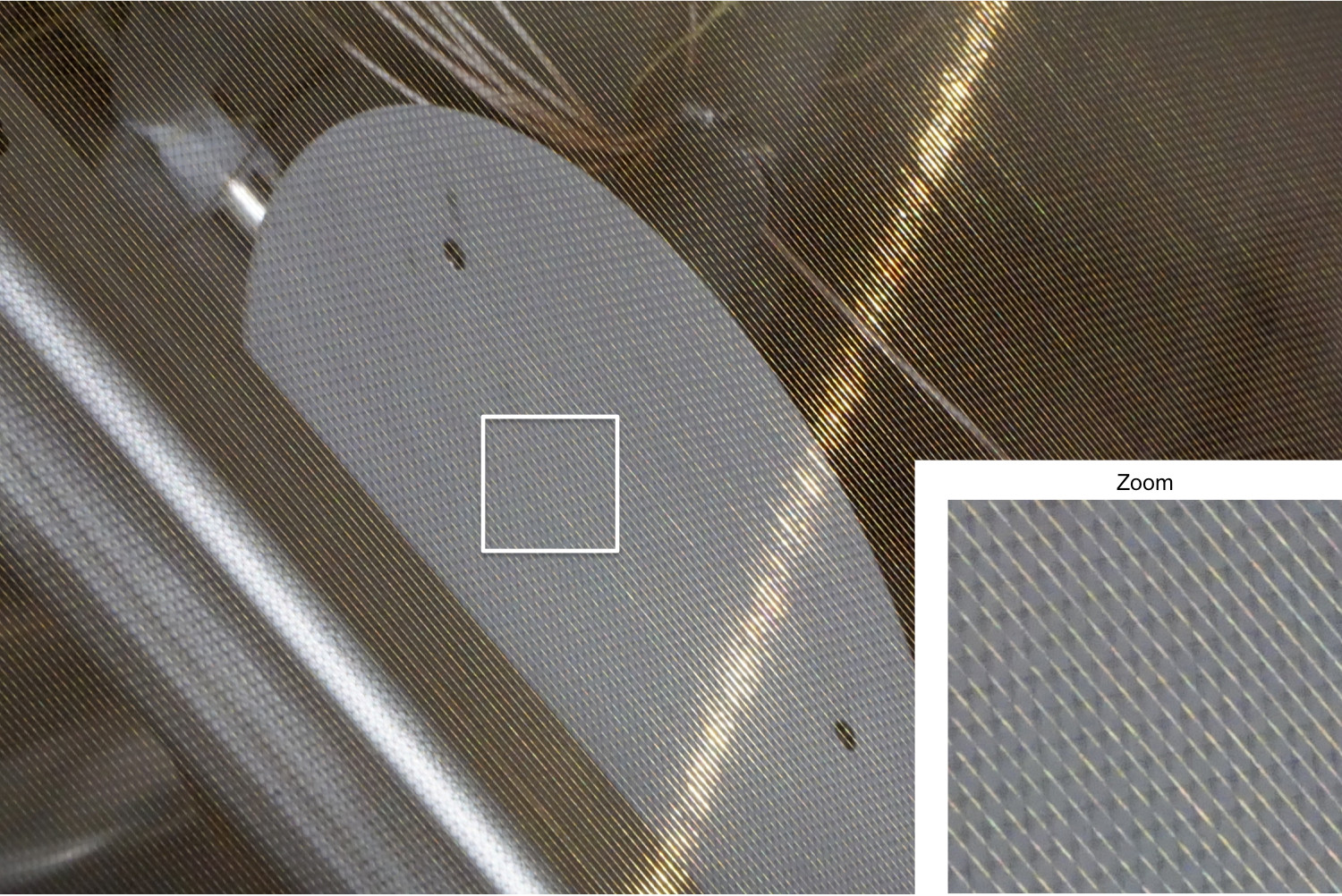}
\caption{View of a central part of the anode plane about halfway between the pump-out port and the corner of the anode plane. The round plate in the background is a wavelength shifting plate covering a PMT. Also visible is the PMT cabling. The picture in the bottom right corner is a zoom of the white square in the large picture.}
\label{fig:410}
\end{figure}

\begin{figure}[H]
\centering
\includegraphics[width=0.87\textwidth]{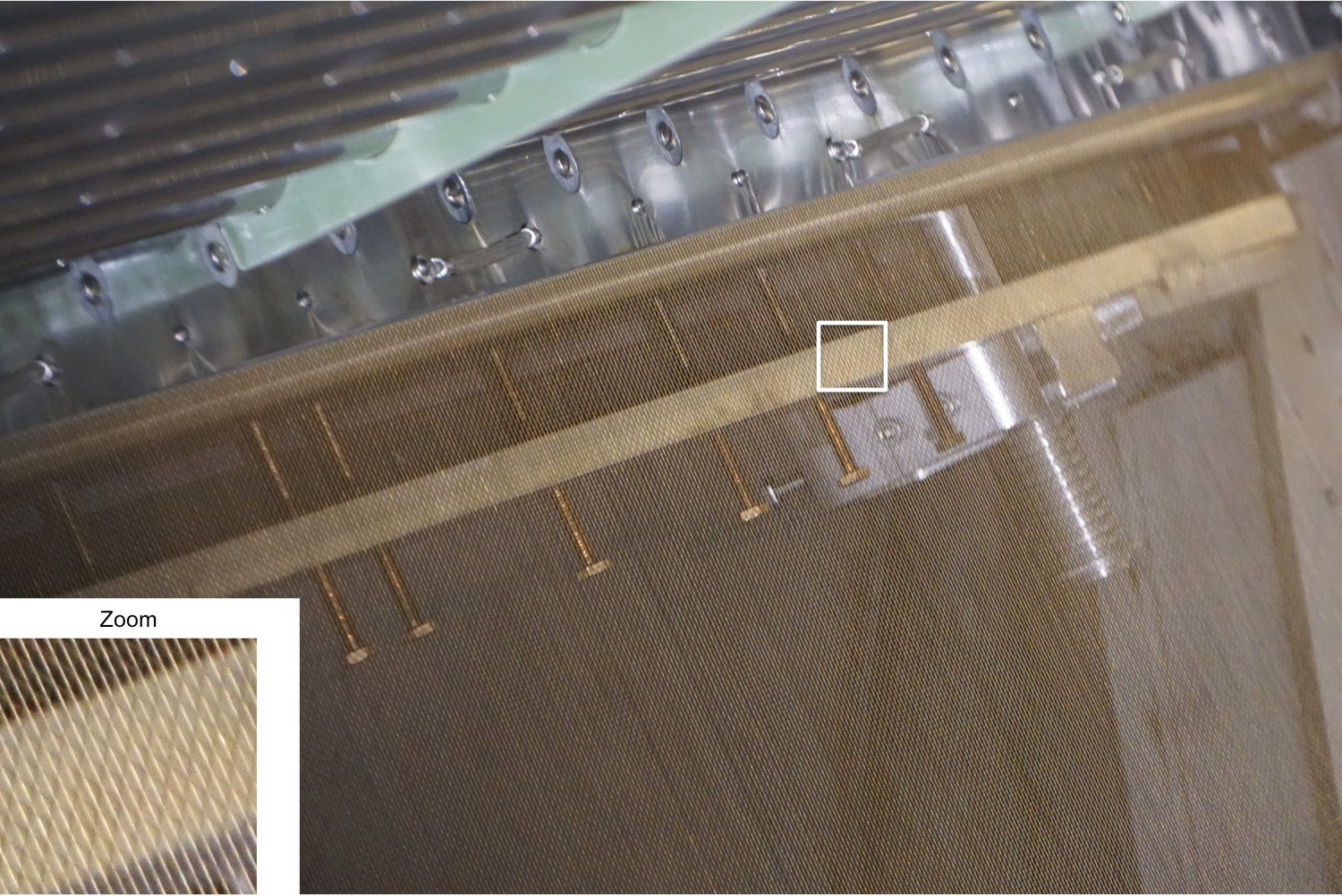}
\caption{View towards the far top corner of the anode wire plane. The distance between the mirror and the wire plane is about 4.20~m. The picture in the bottom left corner is a zoom of the white square in the large picture.}
\label{fig:278}
\end{figure}

All anode areas from the part directly opposite to the pump-out port to the far corner were scanned. This implies that the setup enables the visual inspection of all the TPC anode wires when both pump-out ports are used. Figures~\ref{fig:460} to \ref{fig:278} show some examples of the images taken using the described mirror-camera-light setup (please note that the images have been down-scaled in file size by about 95\% for this report). Multiple pictures were recorded for each mirror rotation, mirror tilt angle, various illumination angles, and various focus settings. The procedure was particularly beneficial for large viewing angles such as towards the corner of the anode wire frame, the largest distance from the mirror. At this maximum distance, the field of view covers a large area of wires which have varying distances to the mirror. Thus, not all wires can be in focus in a single image. The best separation between the wires is naturally obtained for the wires closest to the mirror location. The wires in the corner of the anode plane at the farthest distance can still be resolved to the extent adequate for observing problems, such as possible wire breakage or sagging.\\

The recorded pictures from the interior of the TPC were inspected for any irregularities. The image analysis was performed by multiple people visually scanning all photographs. No distortions or non-uniformities in the wire spacing pattern were seen, indicating no broken or sagging wires within the required MicroBooNE specification of less than 0.5\,mm sag (the resolution of this camera setup resolution was 0.1\,mm over a distance of 5\,m, or smaller for shorter distances).
\\


\section{Conclusion}

With this setup of a commercially available camera, a front surface mirror, and an LED light source, it was demonstrated that the interior of the MicroBooNE cryostat could be inspected. The wire planes, consisting of more than 8,200 gold-plated sense-wires of 150\,\si{\micro\metre} in diameter with a 3~mm spacing, could be resolved from distances of 1.3~m to 5~m. Broken or sagging wires could be excluded from the obtained pictures. The setup was designed to fit through one of the cryogenic service nozzles, which provided the only opportunities to access the cryostat interior. The setup can be easily modified for other cryogenic detectors being used in dark matter or neutrino detection by taking into account their geometrical constraints.\\

\acknowledgments

We would like to acknowledge mechanical support by Patrick Healey and Peter Simon, as well as Keith Anderson for exploring the possibility of using a borescope. We thank Reidar Hahn of Fermilab's Visual Media Services for assistance in selecting and operating the camera. We thank the Fermilab Alignment and Metrology Department for assistance with the calibration of the mirror. We would like to thank the MicroBooNE collaboration for providing test setups and help with the analysis of photographs. Fermilab is operated by the Fermi Research Alliance, LLC under Contract No. DE-AC02-07CH11359 with the United States Department of Energy.

\end{document}